\documentclass[12pt]{article}

\usepackage{scicite}

\usepackage{times}

\usepackage[utf8]{inputenc} 
\usepackage[T1]{fontenc}    
\usepackage[hidelinks]{hyperref}       
\usepackage{url}            
\usepackage{booktabs}       
\usepackage{amsfonts}       
\usepackage{nicefrac}       
\usepackage{microtype}      
\usepackage{lipsum}
\usepackage{graphicx}

\usepackage{amsmath}
\usepackage{calc}
\usepackage{tabularx}

\usepackage{placeins}
\usepackage{flafter}

\usepackage{longtable,booktabs,array}
\usepackage{calc} 
\usepackage{etoolbox}
\makeatletter
\patchcmd\longtable{\par}{\if@noskipsec\mbox{}\fi\par}{}{}
\makeatother
\IfFileExists{footnotehyper.sty}{\usepackage{footnotehyper}}{\usepackage{footnote}}
\makesavenoteenv{longtable}

\newlength{\csllabelwidth}
\newlength{\cslhangindent}
  {}%
  {\par}
\newenvironment{CSLReferences}[2] 
 {
  \setlength{\parindent}{0pt}
  \ifodd #1 \everypar{\setlength{\hangindent}{\cslhangindent}}\ignorespaces\fi
  \ifnum #2 > 0
  \setlength{\parskip}{#2\baselineskip}
  \fi
 }%
 {}
\usepackage{calc} 

\usepackage{mathptmx}
\usepackage[scaled=0.7]{beramono}
\usepackage{xcolor}
\definecolor{crimson}{RGB}{156,0,0}
\usepackage{xurl}
\usepackage[hyphenbreaks]{breakurl}
\hypersetup{colorlinks = true, citecolor = crimson, urlcolor = crimson, linkcolor = crimson}
\usepackage{tocloft}
\usepackage[format=plain, labelfont={bf,it}, textfont=it]{caption}

\usepackage{setspace}

\newenvironment{sciabstract}{%
\begin{quote} \singlespacing}
{\end{quote}}

\newcounter{lastnote}

\title{\bf \textbf{Comment: The Essential Role of Policy Evaluation for the 2020 Census
Disclosure Avoidance System}}

\author{
Christopher T. Kenny,\textsuperscript{1}
Shiro Kuriwaki,\textsuperscript{2}
Cory McCartan,\textsuperscript{3}\\
Evan T. R. Rosenman,\textsuperscript{4}
Tyler Simko,\textsuperscript{1}
and Kosuke Imai\textsuperscript{1, 3}\textsuperscript{*}
\\
\\
\normalsize{\textsuperscript{1}Department of Government, Harvard University}\\
\normalsize{\textsuperscript{2}Department of Political Science, Yale University}\\
\normalsize{\textsuperscript{3}Department of Statistics, Harvard University}\\
\normalsize{\textsuperscript{4}Data Science Initiative, Harvard University}\\
\\
\textsuperscript{*}\small{}To whom correspondence should be addressed. E-mail: \href{mailto:imai@harvard.edu}{\nolinkurl{imai@harvard.edu}}
}

\date{}

\begin{document}

\baselineskip24pt


\maketitle


\begin{sciabstract}
In ``Differential Perspectives: Epistemic Disconnects Surrounding the US Census Bureau's Use of Differential Privacy,'' boyd and Sarathy argue that empirical evaluations of the Census Disclosure Avoidance System (DAS), including our published analysis, failed to recognize how the benchmark data against which the 2020 DAS was evaluated is never a ground truth of population counts.
In this commentary, we explain why policy evaluation, which was the main goal of our analysis, is still meaningful without access to a perfect ground truth.
We also point out that our evaluation leveraged features specific to the decennial Census and redistricting data, such as block-level population invariance under swapping and voter file racial identification, better approximating a comparison with the ground truth.
Lastly, we show that accurate statistical predictions of individual race based on the Bayesian Improved Surname Geocoding, while not a violation of differential privacy, substantially increases the disclosure risk of private information the Census Bureau sought to protect.
We conclude by arguing that policy makers must confront a key trade-off between data utility and privacy protection, and an epistemic disconnect alone is insufficient to explain disagreements between policy choices.
\end{sciabstract}

\begin{center}
Accepted, \emph{Harvard Data Science Review}
\end{center}

\onehalfspacing

\pagebreak

We thank the editors of this special issue for giving us an opportunity to comment on an insightful article, ``Differential Perspectives: Epistemic Disconnects Surrounding the US Census Bureau's Use of Differential Privacy'' by boyd \& Sarathy (2022).
Most academic papers surrounding the use of differential privacy in the 2020 Census have focused on technical issues, including the task of developing a better Census disclosure avoidance system (DAS).
In contrast, boyd and Sarathy considered how the differing perspectives held by policy makers, census officials, and external stakeholders can have significant political impacts.
The authors conducted over 40 interviews to obtain a range of views.\footnote{None of the authors of this commentary was interviewed for the article.}
We agree with boyd and Sarathy that attention to these often-overlooked epistemic issues is also essential.

One of the central claims put forth by boyd and Sarathy is that many researchers and Census users fell prey to the ``statistical illusion'' that the 2010 DAS Census files represent an accurate and objective ``ground truth.'' The authors criticize our work on the DAS and redistricting (Kenny et al., 2021b) by writing that we published our findings ``{[}d{]}espite `ground truth' critiques'' of our earlier working paper (Kenny et al., 2021a).\footnote{boyd and Sarathy relate our working paper with the state of Alabama's litigation against differential privacy, \emph{Alabama v. Commerce}, by writing ``Scientific debates turned into communicative spectacle after lawsuits were filed. First, some external analyses were introduced as expert reports in Alabama v. Commerce {[}\ldots{]} After oral arguments at the district court, a group of political scientists and statisticians released a working paper arguing that the government's system had significant issues that would harm redistricting work.'' Contrary to the perception one might get from this characterization, our working paper was not timed for this lawsuit. Indeed, none of the authors of this article were involved in this lawsuit. We prepared our working paper in response to the Census Bureau's call for feedback based on their April 2021 demonstration data and submitted it just in time for the May 28 deadline (see the acknowledgement section of Kenny et al. (2021a)). Admittedly, some of the language in the working paper was too strong, and we refined it in our published paper (Kenny et al., 2021b) after receiving feedback from other scholars including Bun et al. (2021).}
They contend that we ``ignor{[}ed{]} biases that the published 2010 data might have introduced into the redistricting process.'' Cohen et al. (2022), in the same issue, shares this characterization, writing that the approach in Kenny et al. (2021b) ``ignores the effects of swapping {[}\ldots{]} as well as all of the other documented sources of error known to courts for many years.''

In this commentary, we explain why this ``ground truth'' critique advanced by boyd and Sarathy and others does not diminish the value of our published analysis.
First, we point out that the main goal of our analysis was to empirically evaluate the potential impacts of the Census Bureau's policy change (from the old DAS based on swapping to the new DAS based on differential privacy) on redistricting and its evaluation.
Such \emph{policy evaluation} is meaningful even without the unattainable ``ground truth'' because it sheds light on how a policy change under consideration may influence relevant outcomes.
Our analysis showed that the change of DAS procedure can disproportionately affect certain racial groups and alter the redistricting process and its evaluation.

Second, although the validity of our policy evaluation does not depend on access to the unattainable ``ground truth,'' much of our analysis is based on reasonable approximations to ``ground truth.'' In particular, boyd and Sarathy fail to recognize the fact that the swapping method does not introduce bias to population counts, which represent the key variable used in our redistricting simulation analysis.
In addition, both the election data and the self-reported race in voter files used in our analysis are generally seen as close approximations to their respective ``ground truths.'' Thus, parts of our policy evaluation also speak to the accuracy of redistricting analysis and evaluation to the extent that these data are considered as ``ground truth'' or its reasonable estimate.

Lastly, we show that accurate predictions of individuals' race based on the Bayesian Improved Surname Geocoding (BISG), while not a violation of \emph{differential} privacy, substantially increases the actual disclosure risk of private information the Census Bureau sought to protect when developing the new DAS.
It is important to recognize that differential privacy is one, but not the only, definition of privacy.
Members of the public and policy makers may care about the actual disclosure risk rather than differential privacy.

We conclude this commentary with a discussion of various factors that affect the key trade-off between data utility and privacy protection that is at the heart of the Census DAS debate.
In this regard, we are in complete agreement with Hotz et al. (2022) who point out that insufficient attention has been paid to this trade-off.\footnote{Hotz et al. (2022) use the term ``data usability'' instead of ``data utility'' to refer to the same concept.
  They define it as ``the potential for data to effectively help answer specific questions of interest'' (p.~2).}
More evaluation studies like ours are needed in order to assess the impacts of different disclosure avoidance methods on the data utility of the decennial Census in different policy areas.

\hypertarget{policy-evaluation-is-meaningful-even-without-the-unattainable-ground-truth}{%
\section{Policy evaluation is meaningful even without the unattainable ``ground truth''}\label{policy-evaluation-is-meaningful-even-without-the-unattainable-ground-truth}}

The goal of our analysis published in Kenny et al. (2021b) was to ``empirically evaluate the impact of the DAS, both the noise injection and postprocessing, on redistricting and voting rights analysis across local, state, and federal contexts'' (p.1).
Our study, therefore, should be understood as an example of \emph{policy evaluation} that empirically assesses how a policy change---a change in the Census Bureau's DAS, in this case---might influence outcomes of interest.
Policy evaluation enables policy makers and relevant stakeholders to better understand the potential consequences of a policy change under consideration prior to its implementation.
In fact, partly based on the feedback received from our team and others, the Census Bureau made several important modifications to address concerns about the previous version of the DAS, including those described in Kenny et al. (2021a).\footnote{See the postscript section of Kenny et al. (2021b) for details.}
Regardless of one's opinion about the final DAS, the iterative rounds of feedback solicitation and revision---conducted prior to its final rollout---represent an exemplary policy making process.

Most policy evaluation compares the potential outcomes under one policy (e.g., existing policy) with those of an alternative policy (e.g., new policy).
For this reason, the conceptual and methodological framework of causal inference constitutes the intellectual foundation of policy evaluation (Imbens \& Rubin, 2015).
Our study is no exception.
Specifically, our goal was to compare the potential outcomes of redistricting and voting rights analysis that would result under the old DAS with those under the new DAS.
We do so by applying the same exact analysis procedure to the two April demonstration data sets---one prepared under the old DAS and the other under the new DAS---and comparing the resulting findings.

It is critical to recognize that we are not evaluating the redistricting analysis results under the new DAS against their ``ground truth,'' defined here as the results one would obtain by analyzing the true Census day population counts without any measurement error.
We agree with boyd and Sarathy that such ground truths are never obtainable.
It is well known to social scientists and statisticians that the Census data are at best approximate, and suffer from various undercounts, overcounts, and other types of measurement errors, however small they might be (Anderson \& Fienberg, 1999; Strmic-Pawl et al., 2018).
For example, the Census Bureau itself estimates that the 2020 Census undercounted the Black population by 3.3\% and the Hispanic population by 5\%, while simultaneously overcounting eight entire states (U.S. Census Bureau, 2022).
Thus, the decennial Census is not free of measurement error and does not represent the true population at the time of redistricting.

Nevertheless, many policy makers, analysts and courts have long treated the decennial Census data ``as-is'' while recognizing their potential inaccuracies.
In particular, the Supreme Court wrote in \emph{Karcher v. Daggett} (1983):

\begin{quote}
``The census count provides the only reliable -- albeit less than perfect -- indication of the districts' `real' relative population levels, and furnishes the only basis for good faith attempts to achieve population equality.''
\end{quote}

And yet, in the same case, the Supreme Court also found that the existence of enumeration error is not a justification for deviation from the enumerated population counts.
There exist numerous precedents where the courts relied on relatively small differences in the published P.L.
94-171 population counts to decide whether a district is malapportioned.\footnote{Examples include \emph{Baker v. Carr} (1962), \emph{Reynolds v. Sims} (1964), \emph{Wesberry v. Sanders} (1964), \emph{Kirkpatrick v. Preisler} (1969) and \emph{Anne Arundel County Republican Central Committee v. State Administrative Board of Election Laws} (1991).}
This long-standing practice provides one reason why it is important to empirically evaluate the impacts of the change in the DAS procedure under the assumption that the released Census data would be treated as-is.

We explicitly stated this point in our original paper, writing (p.~1):

\begin{quote}
``Using these demonstration data, we conduct our empirical evaluation under a likely scenario, in which practitioners, map drawers and analysts alike, treat these DAS-protected data''as-is'' as they have done in the past, without accounting for the DAS noise generation mechanism.''
\end{quote}

boyd \& Sarathy (2022) suggest that our focus of ``what would arise `in practice'\,'' should not have been an excuse to analyze the April demonstration data as-is.
To the contrary, we believe that this focus on practice is exactly what is needed when evaluating the impacts of the change in the DAS procedure.
In fact, we are not aware of any redistricting case during this cycle in which map drawers, expert witnesses, or courts have accounted for noise introduced by the new DAS procedure.
All parties involved in these cases treated the 2020 decennial Census as-is, consistent with past practice and what we assumed in our published article.\footnote{In the future, policy makers, expert witnesses, and academic researchers may consider including statistical adjustments for the DAS procedure in their analyses.
  Releasing the so-called ``noisy measurements'' may facilitate such analyses.}

In sum, the unattainable ``ground truth'' is not required for the purpose of evaluating the potential impacts of a policy change.
The main goal of our analysis was not to assess the accuracy of the redistricting analysis results based on the new DAS procedure as compared to those based on unattainable true Census counts.
Instead, we investigated how different the results of redistricting and voting rights analysis might become relative to the current status quo policy once the Census Bureau changed its DAS procedure.
Our analysis showed that this policy change can disproportionately affect certain racial groups and can alter the redistricting process and its evaluation.

\hypertarget{policy-evaluations-can-shed-light-on-accuracy-in-some-cases}{%
\section{Policy evaluations can shed light on accuracy in some cases}\label{policy-evaluations-can-shed-light-on-accuracy-in-some-cases}}

Although the main goal of our study was policy evaluation, which can be conducted without ``ground truth,'' our analysis also provides insights into the accuracy of redistricting and voting analysis based on the new DAS procedure.
In our paper, we demonstrated this in two ways.
First, we accounted for the fact that the confidential data in the Census Edited File --- which, according to boyd and Sarathy, the Census Bureau regards as ground truth\footnote{boyd \& Sarathy (2022) report that ``The Census Bureau knows that its data collection processes are imperfect but believes that, given the operational and confidentiality constraints, the data collected and produced through its well-studied, systematic approach are the best data possible. \emph{Thus, it treats its confidential data as ground truth}.'' (emphasis added)} --- are accurately reflected in public data in terms of marginal population counts.
We took advantage of the population invariance property of the previous DAS procedure (based on the method of swapping), which allowed us to compare the results based on the new DAS procedure against those based on the confidential data without access to such data.
Second, given the potential inaccuracy of the Census confidential data, we also used another potential source of ``ground truth''--- self-reported race from publicly available voter registration files --- to assess the accuracy of racial imputation based on the new DAS procedure.
We explain each of these evaluation strategies below.

\hypertarget{population-invariance-under-swapping}{%
\subsection{Population invariance under swapping}\label{population-invariance-under-swapping}}

A primary contention of boyd \& Sarathy (2022) is that analyses of the new DAS, including ours, incorrectly compared new DAS-protected data with published 2010 Census data, which itself was subject to privacy-protecting distortions in the form of swapping.
Citing Fienberg \& McIntyre (2005), they note that unbeknownst to most external stakeholders, ``techniques like swapping {[}had already{]} introduced distortion into the data''.
As an example, boyd \& Sarathy (2022) claim that we ignored these biases of the 2010 Census data in our published analysis.

As we discuss above, the comparison between swapping-protected and new DAS-protected data is still meaningful for policy evaluation.
But, even if boyd and Sarathy wish to consider whether our analysis is useful as \emph{accuracy evaluation} rather than policy evaluation, their ground truth critique ignores the actual details of the swapping mechanism, leading to an invalid conclusion.
Critically, swapping---unlike the new DAS---holds \emph{invariant} the total population and voting-age population of census blocks.
As the Bureau describes in its documentation of the 2010 SF1 summary file:

\begin{quote}
``A sample of households is selected and matched on a set of selected key variables with households in neighboring geographic areas (geographic areas with a small population) that have similar characteristics (same number of adults, same number of children, etc.).
{[}\ldots{]} there is no effect on the marginal totals for the geographic area with a small population.''
(\href{https://www2.census.gov/programs-surveys/decennial/2010/technical-documentation/complete-tech-docs/summary-file/sf1.pdf\#page=518}{SF1 documentation}, p.~518)
\end{quote}

Since swapped households are matched on the total number of adults and that of children, both total population and voting-age population are preserved under the swapping mechanism.
Citing the same reasoning, Census Bureau statisticians also compared the TopDown algorithm to the public swapped dataset in their report studying population parity (Wright \& Irimata, 2021, p. 2).

Thus, all of the main analyses in our paper about population parity and ``One Person One Vote'' (Kenny et al. (2021b), Figs. 3-4), and similar analyses by other authors, actually compare DAS-protected data to what is treated by the Census Bureau as ``ground truth'' (see footnote 7).
In addition, our partisan gerrymandering analysis (Kenny et al. (2021b), Fig. 5) uses only total population data to perform the redistricting simulation, and election results to tabulate district vote shares.
These official election results are reported exactly by state and county election boards and are not subject to privacy protection.\footnote{Of course, the official election data may also contain small counting errors and hence, like the Census Edited File, are not necessarily ``ground truth'' in a strict sense.}
Here, too, the comparison is between an analysis of the new DAS-protected data and the same analysis of what the Census Bureau considers ``ground truth'' data.

While the swapping method used in the past decade preserves total and voting-age population, the new DAS procedure does not-----it only preserves the total number of households (and the total population at the state level).
As we showed in our paper, this difference can have a large effect on the meaning of ``One Person One Vote,'' as defined by courts and implemented by policy makers, inflating nominal population balance thresholds by five times or more.
Our partisan analysis also demonstrates the potential impacts of the change in DAS procedure on the evaluation of partisan gerrymandering.

While swapping adds some statistical noise to characteristics of neighboring geographic areas, its block-level population invariance property is powerful.
Under the One Person One Vote principle, plans with a population deviation above certain thresholds can be presumed to be unconstitutional (see \emph{White v. Regester} (1973) or \emph{Karcher v. Daggett} (1983) for some examples).
Since block populations are no longer invariant under the new DAS, the universe of constitutional plans is different from the one that would be obtained under the old DAS.
Depending on random noise injected by the new DAS, plans which are presumptively constitutional can lose that presumption, while other plans may gain that presumption.
Although boyd \& Sarathy (2022) use Cohen et al. (2021) to rebut some of our original claims, both articles fail to appreciate the fact that actual redistricting can be sensitive to minor perturbations or inaccuracies in these numbers.\footnote{Authors of Cohen et al. (2021) appear alongside boyd \& Sarathy (2022) in the HDSR special issue as Cohen et al. (2022) and make similar claims.
  Both Cohen et al. (2021) and Cohen et al. (2022) do not acknowledge the possibility that because DAS noise is applied to the Census Edited File, courts may decide that DAS noise is qualitatively different from previously known sources of error already included in the Census Edited File.
  This qualitatively different noise perturbs the populations in the released data, which changes the space of legal plans away from those possible under the Census Edited File.}
If noise added to population counts were completely random, the space of possible plans \emph{would} change, but this change would be generally uncorrelated with outcomes like race or partisanship.
As we showed in Kenny et al. (2021b), however, these errors were strongly correlated with racial diversity in the April 2020 DAS release, leading to unintended consequences in terms of both race and partisanship.

Beyond redistricting, census population counts are the sole or primary determinant of the apportionment and distribution of federal funds to states, counties, and municipalities.
Addition of noise to population totals must therefore be done carefully due to its more central role in funding apportionment than many other published Census statistics.
Our analysis shows that the new DAS concentrated these population errors disproportionately on minority groups and Democratic voters, suggesting potential disparate impacts of the change in the DAS procedure.\footnote{Our findings do not necessarily imply that swapping should be preferred over the differential privacy in general.
  For example, it is possible to develop a differential privacy procedure that preserves population counts (e.g., Gong \& Meng, 2020).}

\hypertarget{the-use-of-self-reported-race-in-voter-files-as-ground-truth}{%
\subsection{The use of self-reported race in voter files as ``ground truth''}\label{the-use-of-self-reported-race-in-voter-files-as-ground-truth}}

boyd and Sarathy's ``illusion of ground truth'' critique also does not apply to our accuracy evaluation of racial imputation based on Bayesian Improved Surname Geocoding (BISG).
In that analysis, we relied upon the self-reported race of individual voters in North Carolina's voter registration records, which are publicly available on the Internet for download.
To the extent that an individual's self-reported race can be considered ``ground truth,'' our analysis showed that the new DAS procedure does not negatively impact BISG's predictive accuracy when compared to the old DAS procedure.

This finding is notable because the Census Bureau's primary motivation for adopting the new DAS procedure was the protection of individuals' self-reported race in the confidential file (Abowd, 2021).
Thus, while the moderate amount of noise added by the DAS negatively impacted the utility of the 2020 Census data for redistricting purposes, the DAS left unchanged the ability to accurately \emph{infer} individual's race, which they sought to protect.

\hypertarget{accurate-statistical-prediction-increases-individual-disclosure-risk}{%
\subsection{Accurate statistical prediction increases individual disclosure risk}\label{accurate-statistical-prediction-increases-individual-disclosure-risk}}

The initial version of our working paper (Kenny et al., 2021a), as it was submitted to the Census Bureau for public comment, attracted critiques by other scholars.
Most notably, a group of differential privacy scholars posted a commentary (Bun et al., 2021) on the interpretation of our findings regarding the use of BISG to predict individual-level race.
In boyd and Sarathy's accounting of this exchange, our working paper ``incensed other Harvard researchers, some of whom issued a technical rebuttal highlighting how the comparison data was not ground truth.''

It is important to note that boyd and Sarathy mischaracterize Bun et al.'s critique.
Bun et al. (2021) are not occupied by whether the comparison data represent ``ground truth.'' In fact, there is no reason to believe that self-reported individual race in voter files is less accurate than the corresponding information in the Census confidential data.
Rather, Bun et al. (2021) take issue with our interpretation of the probabilistic prediction of individual race based on the DAS data.
They contend that our BISG exercise is that of statistical inference and prediction, so its success in predicting individual race is not a privacy violation.

We agree with Bun et al. (2021) that accurate BISG predictions do not constitute a violation of \emph{differential} privacy (Kenny et al., 2021b, p. 13). Differential privacy is a mathematical definition of privacy which provides a probabilistic guarantee on a certain type of disclosure risk. We do not dispute that the new Census DAS builds upon this formal definition, but differential privacy is not the only definition of privacy. Privacy is a concept that has several legal, technical, and non-technical definitions, and the courts, general public, and policy makers may not agree with differential privacy (Cormode et al., 2013; e.g., Hotz et al., 2022; Hotz \& Salvo, 2022; Rubel, 2011; Solove, 2002).

Indeed, a primary goal of the Census Bureau in adopting the new DAS was to prevent inference of individuals' ``private'' racial information (Abowd, 2019). This notion of disclosure is close to the operationalization of privacy as \emph{absolute disclosure risk} rather than the DP criterion (Duncan \& Lambert, 1986, 1989).
The absolute disclosure risk is defined as the ``probability that an `intruder' can identify individuals in a data set, based on the released data, {[}and{]} what intruders know about individuals in the absence of such data'' (Hotz \& Salvo, 2022).
Highly accurate statistical prediction of individual race is just as much a violation of this notion of privacy as a reidentification accomplished through database reconstruction.\footnote{Like Hotz et al. (2022), we are aware of the difficulty of controlling the absolute disclosure (Dwork \& Naor, 2010).
  Nevertheless, the multifaceted nature of privacy means that members of the public and policymakers may feel that using Census data to accurately predict their racial information is a violation of their privacy even though the new DAS protects one particular definition of privacy.}
This point was also made in a 2022 report commissioned by the Bureau itself (JASON, 2022) (p.~114):

\begin{quote}
``In the context of the decennial census, there is no risk to an individual if it is learned that their data are included in the census.
The decennial data includes all United States persons whether or not they self-report data to the Census Bureau\ldots.
\emph{Hence, the concern for the Census Bureau should be focused on inferential disclosure risk} {[}emphasis added{]}.
The risk that matters is if the released data allows an adversary to make inferences about an individual's characteristics with more accuracy and confidence than could be done without the data released by the Census Bureau.''
\end{quote}

Specifically, the absolute disclosure risk is defined as \(\Pr(J=j, Y_j\mid D^*, A)\), where \(A\) is an attacker's dataset, \(D^*\) is a differentially private data release, \(Y\) is the protected attribute, \(J\) is the target individual in \(A\), and \(j\) is an individual in \(D^*\).
Hotz et al. (2022) assume that \(D^*\) takes the form of individual records with \(Y_j\) observed.
In that context, \((J=j, Y_j)\) means that a particular entry \(Y_j\) in \(D^*\) is associated to the target individual \(J\).
In the case of decennial Census, every individual in the U.S. is supposed to be in the data, but \(D^*\) consists of noised tables for each Census geography rather than microdata.
Therefore, \((J=j, Y_j)\) is equivalent to identification of the unobserved attribute \(Y_J\) for the target individual, i.e., \(\Pr(J=j, Y_j \mid D^\ast, A)=\Pr(Y_J\mid D^*, A)\).\footnote{Since the residence location of each registered voter is publicly available as part of voter files, inferring their race is equivalent of identifying them in a census table of race counts.}
This is the same connection Hotz et al. (2022) make between their notation and that of Gong \& Meng (2020) who refer to the absolute disclosure risk as the \emph{actual posterior risk of disclosure} and write it as \(\Pr(Y_J\mid D^*)\), where the conditioning on auxiliary data \(A\) is absorbed into the prior.
This is precisely what BISG produces: the probability that a target individual \(J\) who is in the Census data has a particular self-reported race (\(Y_J\)), conditioning on the information contained in the Census data release (\(D^*\)) and using auxiliary name data (\(A\)) as a prior.

Of course, one must take into account the baseline disclosure risk when assessing the role of a particular data release in contributing to the absolute risk of disclosure.
As in Kenny et al. (2021b), we evaluate the North Carolina voter file obtained in February 2021 through L2 Inc., a leading national nonpartisan firm that supplies voter data and related technology.
Unsurprisingly, we find that the decennial Census, even under the new DAS, substantially increases the absolute disclosure risk, compared to a baseline without any detailed Census data.
Table \ref{tab:bisg-error} shows the error rate in BISG predictions when using 2010 Census data, as well as the error rates when not using the data (i.e., only using name information to predict race), for the North Carolina data.
Misclassification rates are computed by assigning every individual to the maximum a posteriori class and comparing against true, self-reported race.
The results show that the inclusion of detailed Census geographic and racial data substantially improves the prediction accuracy and thus the absolute risk of disclosure.

The last column presents the maximum observed relative risk of disclosure for an individual in the dataset: the factor by which our confidence in their individual racial information increased or decreased going from the name-only predictions to the name-and-Census-data predictions.
This relative risk is precisely the quantity controlled by the differential privacy guarantee (Gong \& Meng, 2020).
The numbers here should be considered as a lower bound; they could be higher depending on the presence of other auxiliary information.
While the relative disclosure risk is quite large for some individuals, it is not a violation of differential privacy---the guarantees provided by the large privacy loss budget of \(\varepsilon=19.61\) in the new DAS are extremely weak.\footnote{Gong \& Meng (2020) show that \(\varepsilon\) can be interpreted as a bound on how much the prior probability of an individual having a certain protected attribute can change under Bayesian updating from differentially private data release.
  Specifically, the prior probabilities can change by no more than a factor of \(\exp(\pm\varepsilon)\).
  With \(\varepsilon=19.61\), this factor is over 328 \emph{million}.
  As the Bureau-commissioned report writes, ``the formal guarantees provided by differential privacy at this level provide little guaranteed uncertainty'' (JASON, 2022, p. 67).}
Nevertheless, the substantially higher absolute risk of disclosure indicated by the lower error rates, as well as the large relative risk, could potentially be concerning to members of the public who are worried about the disclosure of racial self-identification.
Moreover, these risks are not borne equally: the \emph{average} individual relative risk varies substantially by race, from 1.96 for White voters to as high as 14.0 and 21.5 for Asian and ``Other'' voters, respectively.\footnote{These averages were calculated as the geometric mean of the individual relative risk for BISG using first, middle, and last names.
  The lower average risk for White voters stems from the correspondingly high prior probability of being white, given U.S. demographics.
  The ``Other'' category includes those selecting two or more races as well as the American Indian / Alaska Native population.}

\begin{longtable}[]{@{}
  >{\raggedright\arraybackslash}p{(\columnwidth - 6\tabcolsep) * \real{0.3647}}
  >{\raggedright\arraybackslash}p{(\columnwidth - 6\tabcolsep) * \real{0.2118}}
  >{\raggedright\arraybackslash}p{(\columnwidth - 6\tabcolsep) * \real{0.2118}}
  >{\raggedright\arraybackslash}p{(\columnwidth - 6\tabcolsep) * \real{0.2118}}@{}}
\caption{\label{tab:bisg-error} BISG error rates based on the data sources used in racial predictions. All examples use the overall surname-by-race table also published by the Census Bureau. Maximum individual relative disclosure risk is calculated as the maximum possible ratio \(\Pr(Y_J|D^*)/\Pr(Y_J)\) (or its inverse) with \(Y_J\) the race of individual \(J\) and \(D^*\) the DAS-19.61 data, with the prior \(\Pr(Y_J)\) representing the name-only predictions.}\tabularnewline
\toprule()
\begin{minipage}[b]{\linewidth}\raggedright
BISG Method
\end{minipage} & \begin{minipage}[b]{\linewidth}\raggedright
Error rate without Census data
\end{minipage} & \begin{minipage}[b]{\linewidth}\raggedright
Error rate with DAS-19.61 data
\end{minipage} & \begin{minipage}[b]{\linewidth}\raggedright
Maximum individual relative disclosure risk
\end{minipage} \\
\midrule()
\endfirsthead
\toprule()
\begin{minipage}[b]{\linewidth}\raggedright
BISG Method
\end{minipage} & \begin{minipage}[b]{\linewidth}\raggedright
Error rate without Census data
\end{minipage} & \begin{minipage}[b]{\linewidth}\raggedright
Error rate with DAS-19.61 data
\end{minipage} & \begin{minipage}[b]{\linewidth}\raggedright
Maximum individual relative disclosure risk
\end{minipage} \\
\midrule()
\endhead
Only last names & 40.9\% & 15.5\% & 796.9 \\
First and last names & 27.5\% & 12.4\% & 969.6 \\
First, middle, and last names & 19.0\% & 10.2\% & 1077.8 \\
\bottomrule()
\end{longtable}

boyd \& Sarathy (2022) argue that those who embrace differential privacy harness uncertainty, while those who reject differential privacy ``view centering uncertainty as politically, legally, and rhetorically dangerous; they prefer statistical illusions, however technically imperfect.'' Such a characterization of the Census DAS debate is overly simplistic, because other definitions of privacy, including absolute risk of disclosure presented here, incorporate uncertainty as well.

In sum, although accurate statistical prediction is not a violation of differential privacy, a highly accurate probabilistic assessment of individuals' private information can be problematic from other privacy protection points of view.
Members of the public and policy makers may care about other notions of privacy such as absolute risk of disclosure.
As our analysis shows, the release of decennial Census under the new DAS can still substantially increase the accuracy of these probabilistic assessments.
A similar point is made in the literature on fairness in decision-making.
For example, even if race is not directly used in a decision, an accurate proxy of race can be leveraged to discriminate against individuals of certain racial groups.
Our analysis addressed the question of how the change in DAS affects the accuracy of predicting the private information the Census Bureau sought to protect.

\hypertarget{concluding-remarks-trade-off-between-data-utility-and-privacy-protection}{%
\section{Concluding remarks: trade-off between data utility and privacy protection}\label{concluding-remarks-trade-off-between-data-utility-and-privacy-protection}}

boyd \& Sarathy (2022) made a valuable contribution by drawing attention to the epistemic issues surrounding the 2020 Census DAS. We agree with the authors that the trust relationships between the Census Bureau and Census data users play an essential role in the credibility of the Census data among policy makers and researchers.
In this commentary, we responded to the ``illusion of ground truth'' critique offered by boyd and Sarathy against our published work.
We argued that policy evaluation conducted in Kenny et al. (2021b) is important even if no agreement exists about what constitutes ``ground truth.'' Evaluating the change in outcomes from the status quo (old DAS with swapping) to a new policy (new DAS with differential privacy) is a meaningful exercise because the current legal and practical structure of redistricting uses Census data as-is.
We also showed that many of our analyses can illuminate the accuracy of redistricting and voting rights analysis by exploiting the availability of certain data and their properties.
Lastly, we pointed out that differential privacy is only one definition of privacy and other conceptualizations of privacy may also play an important role when developing an optimal DAS procedure.
For example, accurate statistical predictions of individuals' private information can lead to the increased disclosure risk, which members of the public and policy makers may find problematic.

The Census Bureau is bound by law to protect respondent privacy.
Our analysis presented in Kenny et al. (2021b) does not question this goal.
Policy makers, however, must balance between data utility and privacy protection.
This consideration is essential given the paramount importance of the decennial Census in policy making.
Even if the Census Bureau, the public, and other external stakeholders agree on differential privacy as an appropriate privacy framework, the right amount of noise to inject should be informed by the specific privacy harms the Bureau wants to protect against, as well as the ``base rate'' accuracy of statistical inference of protected attributes (Francis, 2022).

The policy evaluation in Kenny et al. (2021b) examines this key tradeoff that is inherent in the DAS.
We demonstrated that the moderate amount of error injected by the DAS degrades the utility of the Census data for redistricting purposes while failing to reduce the ability to accurately infer the individual race information that the Census Bureau claims it needs to protect.
Hotz et al. (2022) point out that ``the discussion {[}surrounding the DAS{]} has focused almost entirely on privacy protection, with little assessment of the impact on data usability and, hence, social knowledge'' (p.~9).
Our study helped fill this gap by examining how this change in the DAS affects data usability in legislative redistricting.
The Bureau-commissioned study (JASON, 2022) reinforces the need for these kinds of studies:

\begin{quote}
``The Census Bureau sought to satisfy utility needs while minimizing formal privacy loss, but concrete disclosure risks are not sufficiently quantified to factor into decisions about disclosure avoidance options. {[}\ldots{]}
It is unclear if the privacy mechanisms adopted are sufficient to mitigate the vulnerabilities.''
(pp.~4-5)
\end{quote}

The data used for our evaluation of BISG prediction raises an additional policy consideration when weighing the cost and benefit of the change in the DAS procedure.
In particular, individual self-reported race, which the Census Bureau sought to protect, is already publicly available for all registered voters in several southern states (Alabama, Florida, Georgia, Louisiana, North Carolina, and South Carolina) in their voter files.
Although selection into the voter file is a voluntary choice, about 40 million Americans' self-reported race is available in these public records.
When weighing the trade-off between data utility and privacy protection, it is important to consider what information is already available, what constitutes private information, and whether such private information can already be accurately predicted.

The 2010 and 2020 Census may comprise a ``statistical imaginary,'' in which different stakeholders possess different understandings and use cases for data released by the Census Bureau and their utility.
But, policy changes can have meaningful real-world impacts even if they incorporate imperfect measures of their goals.
Although our focus was redistricting, this is an important feature of any policy change.
For example, a change in the question format of standardized tests may yield disparate impacts among different groups of students even if these tests do not perfectly measure student readiness for college.
Our original article documented how the proposed change in the Census DAS procedure would affect the usability of the data.

\hypertarget{disclosure-statement}{%
\subsection*{Disclosure Statement}\label{disclosure-statement}}
\addcontentsline{toc}{subsection}{Disclosure Statement}

The authors have no disclosures to share for this manuscript.

\hypertarget{acknowledgements}{%
\section*{Acknowledgements}\label{acknowledgements}}
\addcontentsline{toc}{section}{Acknowledgements}

We thank Georgie Evans for her helpful comments on an earlier version of this paper.

\newpage

\hypertarget{references}{%
\section*{References}\label{references}}
\addcontentsline{toc}{section}{References}

\hypertarget{refs}{}
\begin{CSLReferences}{1}{0}
\leavevmode\vadjust pre{\hypertarget{ref-abowdpres2019}{}}%
Abowd, J. M. (2019). \emph{Staring down the database reconstruction theorem}. \url{https://perma.cc/L25P-6EBL}

\leavevmode\vadjust pre{\hypertarget{ref-abowd-reconstruct}{}}%
Abowd, J. M. (2021). \emph{{Affidavit Declaration of John M. Abowd}}. Defendants' response in opposition to combined motion for a preliminary injunction and petition for a writ of mandamus, \emph{Alabama v. U.S. Department of Commerce}, No. 3: 21-cv-211-RAH-KFP. \url{https://perma.cc/E6TJ-SSTU}

\leavevmode\vadjust pre{\hypertarget{ref-censuscount}{}}%
Anderson, M., \& Fienberg, S. E. (1999). \emph{Who counts?: The politics of census-taking in contemporary america}. Russell Sage Foundation.

\leavevmode\vadjust pre{\hypertarget{ref-boydSarathy}{}}%
boyd, danah, \& Sarathy, J. (2022). Differential perspectives: Epistemic disconnects surrounding the US census bureau's use of differential privacy. \emph{Harvard Data Science Review}. \url{https://doi.org/10.1162/99608f92.66882f0e}

\leavevmode\vadjust pre{\hypertarget{ref-buncomment}{}}%
Bun, M., Desfontaines, D., Dwork, C., Naor, M., Nissim, K., Roth, A., Smith, A., Steinke, T., Ullmanand, J., \& Vadhan, S. (2021). \emph{Statistical inference is not a privacy violation}. \url{https://differentialprivacy.org/inference-is-not-a-privacy-violation}

\leavevmode\vadjust pre{\hypertarget{ref-Cohen2021FORC}{}}%
Cohen, A., Duchin, M., Matthews, J. N., \& Suwal, B. (2021). \emph{Census TopDown: The impacts of differential privacy on redistricting}. \url{https://doi.org/10.4230/LIPICS.FORC.2021.5}

\leavevmode\vadjust pre{\hypertarget{ref-Cohen2022Private}{}}%
Cohen, A., Duchin, M., Matthews, J. N., \& Suwal, B. (2022). Private {Numbers} in {Public} {Policy}: Census, {Differential} {Privacy}, and {Redistricting}. \emph{Harvard Data Science Review}, \emph{Special Issue 2}. \url{https://doi.org/10.1162/99608f92.22fd8a0e}

\leavevmode\vadjust pre{\hypertarget{ref-cormode2013}{}}%
Cormode, G., Procopiuc, C. M., Shen, E., Srivastava, D., \& Yu, T. (2013). Empirical privacy and empirical utility of anonymized data. \emph{2013 IEEE 29th International Conference on Data Engineering Workshops (ICDEW)}, 77--82. \url{https://doi.org/10.1109/ICDEW.2013.6547431}

\leavevmode\vadjust pre{\hypertarget{ref-duncan1986disclosure}{}}%
Duncan, G., \& Lambert, D. (1986). Disclosure-limited data dissemination. \emph{Journal of the American Statistical Association}, \emph{81}(393), 10--18. \url{https://doi.org/10.1080/01621459.1986.10478229}

\leavevmode\vadjust pre{\hypertarget{ref-duncan1989risk}{}}%
Duncan, G., \& Lambert, D. (1989). The risk of disclosure for microdata. \emph{Journal of Business \& Economic Statistics}, \emph{7}(2), 207--217. \url{https://doi.org/10.2307/1391438}

\leavevmode\vadjust pre{\hypertarget{ref-dwork2010difficulties}{}}%
Dwork, C., \& Naor, M. (2010). On the difficulties of disclosure prevention in statistical databases or the case for differential privacy. \emph{Journal of Privacy and Confidentiality}, \emph{2}(1). \url{https://doi.org/10.29012/jpc.v2i1.585}

\leavevmode\vadjust pre{\hypertarget{ref-francis2022note}{}}%
Francis, P. (2022). A note on the misinterpretation of the US census re-identification attack. \emph{Proceedings of the International Conference Privacy in Statistical Databases}. \url{https://doi.org/10.1007/978-3-031-13945-1_21}

\leavevmode\vadjust pre{\hypertarget{ref-gong2020congenial}{}}%
Gong, R., \& Meng, X.-L. (2020). Congenial differential privacy under mandated disclosure. \emph{Proceedings of the 2020 ACM-IMS on Foundations of Data Science Conference}, 59--70. \url{https://doi.org/10.1145/3412815.3416892}

\leavevmode\vadjust pre{\hypertarget{ref-hotz_pnas}{}}%
Hotz, V. J., Bollinger, C. R., Komarova, T., Manski, C. F., Moffitt, R. A., Nekipelov, D., Sojourner, A., \& Spencer, B. D. (2022). Balancing data privacy and usability in the federal statistical system. \emph{Proceedings of the National Academy of Sciences}, \emph{119}(31), e2104906119. \url{https://doi.org/10.1073/pnas.2104906119}

\leavevmode\vadjust pre{\hypertarget{ref-Hotz2022Chronicle}{}}%
Hotz, V. J., \& Salvo, J. (2022). A {Chronicle} of the {Application} of {Differential} {Privacy} to the 2020 {Census}. \emph{Harvard Data Science Review}, \emph{Special Issue 2}. \url{https://doi.org/10.1162/99608f92.ff891fe5}

\leavevmode\vadjust pre{\hypertarget{ref-imbensrubin}{}}%
Imbens, G. W., \& Rubin, D. B. (2015). \emph{{Causal Inference for Statistics, Social, and Biomedical Sciences: An Introduction}}. Cambridge University Press. \url{https://doi.org/10.1017/cbo9781139025751}

\leavevmode\vadjust pre{\hypertarget{ref-jason2022}{}}%
JASON. (2022). \emph{Consistency of data products and formal privacy methods for the 2020 {Census} (JSR-21-02, {January} 11, 2022)}. The {MITRE} Corporation. \url{https://perma.cc/XJS8-ADX6}

\leavevmode\vadjust pre{\hypertarget{ref-kenny2021working}{}}%
Kenny, C. T., Kuriwaki, S., McCartan, C., Rosenman, E. T. R., Simko, T., \& Imai, K. (2021a). The impact of the {U.S.} Census disclosure avoidance system on redistricting and voting rights analysis. \emph{arXiv}. \url{https://doi.org/10.48550/arXiv.2105.14197}

\leavevmode\vadjust pre{\hypertarget{ref-kenny2021}{}}%
Kenny, C. T., Kuriwaki, S., McCartan, C., Rosenman, E. T., Simko, T., \& Imai, K. (2021b). The use of differential privacy for census data and its impact on redistricting: The case of the 2020 US census. \emph{Science Advances}, \emph{7}(41), eabk3283. \url{https://doi.org/10.1126/sciadv.abk3283}

\leavevmode\vadjust pre{\hypertarget{ref-rubel2011particularized}{}}%
Rubel, A. (2011). The particularized judgment account of privacy. \emph{Res Publica}, \emph{17}(3), 275--290. \url{https://doi.org/10.1007/s11158-011-9160-4}

\leavevmode\vadjust pre{\hypertarget{ref-solove2002conceptualizing}{}}%
Solove, D. J. (2002). Conceptualizing privacy. \emph{California Law Review}, 1087--1155. \url{https://doi.org/10.2307/3481326}

\leavevmode\vadjust pre{\hypertarget{ref-racecounts}{}}%
Strmic-Pawl, H. V., Jackson, B. A., \& Garner, S. (2018). Race counts: Racial and ethnic data on the {U.S.} Census and the implications for tracking inequality. \emph{Sociology of Race and Ethnicity}, \emph{4}(1), 1--13. \url{https://doi.org/10.1177/2332649217742869}

\leavevmode\vadjust pre{\hypertarget{ref-census2022undercount}{}}%
U.S. Census Bureau. (2022). Census bureau releases estimates of undercount and overcount in the 2020 census. \emph{March 10, 2022, Release Number {CB22-CN.02}}. \url{https://perma.cc/9XBQ-ECJN}

\leavevmode\vadjust pre{\hypertarget{ref-wright2021census}{}}%
Wright, K., \& Irimata, K. (2021). Empirical study of two aspects of the topdown algorithm output for redistricting: Reliability \& variability (august 5, 2021 update). \emph{{U.S. Census Bureau Study Series}}, \emph{Statistics 2021-02}. \url{https://perma.cc/9EPK-P4G6}

\end{CSLReferences}

\bibliography{references.bib}
\bibliographystyle{Science}

\renewcommand{\thetable}{\arabic{table}}
\renewcommand{\thefigure}{\arabic{figure}}
\setcounter{table}{0}
\setcounter{figure}{0}










\end{document}